\documentclass[journal,12pt,onecolumn,draftclsnofoot,]{IEEEtran}

\usepackage{graphicx}
\usepackage{float} 
\usepackage{amssymb}

\usepackage{lineno}
\usepackage{xcolor}  
\usepackage{siunitx} 
\usepackage{svrsymbols} 
\usepackage[utf8]{inputenc} 
\usepackage[T1]{fontenc} 

\usepackage{cite}

\ifCLASSINFOpdf
\else
\fi
\begin{document}

\title{Development of a High Rate Front-end ASIC for X-ray Spectroscopy and Diffraction Applications}
%
%
%


\author{Emerson~Vernon,
        Gianluigi~De~Geronimo,~\IEEEmembership{}
        Jonathan~Baldwin,
        Wei~Chen,
        Jack~Fried,
        Gabriele~Giacomini,
        Anthony~Kuczewski,
        John~Kuczewski,
        Joe~Mead,
        Antonino~Miceli,
        John~S.~Okasinski,
        Don~Pinelli,
        Orlando~Quaranta,
        Abdul~K.~Rumaiz,
        Peter~Siddons,
        Graham~Smith,
        Milutin~Stanacevic,
        and~Russell~Woods~\IEEEmembership{}

\thanks{E. Vernon, W. Chen, J. Fried, G. Giacomini, J. Kuczewski, J. Mead, D. Pinelli, and G. Smith are with the Instrumentation
Division, Brookhaven National Laboratory, Upton, NY 11973
USA e-mail: (evernon@bnl.gov;  weichen@bnl.gov; jfried@bnl.gov; giacomini@bnl.gov; jkuczewski@bnl.gov; mead@bnl.gov; pinelli@bnl.gov; gsmith@bnl.gov).}

\thanks{E. Vernon, G. De Geronimo, and M. Stanacevic  are with the Department of Electrical \& Computer Engineering, Stony Brook University, NY 11794
USA e-mail: (emerson.vernon@stonybrook.edu; degeronimo@ieee.org; milutin.stanacevic@stonybrook.edu).}

\thanks{A. Kuczewski, A. K. Rumaiz, and P. Siddons are with the NSLS II, Brookhaven National Laboratory, Upton, NY 11973
USA email: (kuczewski@bnl.gov; rumaiz@bnl.gov; siddons@bnl.gov).}

\thanks{J. Baldwin, A. Miceli, J. Okasinski, O. Quaranta, and R. Woods are with the X-ray Science Division, Argonne National Laboratory, IL 60439
USA email: (jbaldwin@anl.gov; amiceli@anl.gov; okasinski@anl.gov; oquaranta@anl.gov; rwoods@anl.gov).}

\thanks{Manuscript received \today}}
\maketitle


\begin{abstract}
We developed a new front-end applicatin specific integrated circuit (ASIC) for the upgrade of the Maia x-ray microprobe. The ASIC instruments 32 configurable front-end channels that perform either positive or negative charge amplification, pulse shaping, peak amplitude and time extraction along with buffered analog storage. At a gain of 3.6 V/fC, 1 \si{\micro\second} peaking time and a temperature of 248 K, an electronic resolution of 13- and 10 \si{\electron} rms was measured at  with and without a  SDD sensor respectively. A spectral resolution of 170 eV FWHM at 5.9 keV was obtained with an $^{55}$Fe source. The channel linearity was better than $\pm$ 1 \si{\percent} with rate capabilities up to 40 kcps. The ASIC was fabricated in a commercial 250 nm process with a footprint of 6.3 mm x 3.9 mm and dissipates 167 mW of static power.  
\end{abstract}


%
\IEEEpeerreviewmaketitle


\section{Introduction}
\label{Introduction}
%
%
%
%

\IEEEPARstart{C}{ost} effective and non-destructive elemental mapping of complex material matrices with minimal sample preparation have attracted significant interest across a wide range of disciplines and industries. As a result, high resolution elemental x-ray analysis has become one of the mainstream analytical tools used in exploration, industrial production, pharmaceuticals, forensics, arts and archaeology where quality, safety, and authenticity are of high importance \cite{Arzhantsev:2011ac,Shackley:2011m,Mantler:2000xrs,Bonvin:2006eac}. 

Advances in x-ray sensors and application-specific integrated circuits (ASICs) have played an integral role in the deployment of this technique\cite{Gatti:1984nim,Maniguet:2012iop,Newbury:2013jas,DeGeronimo:2010tns}. Maia, a compact multichannel x-ray microprobe with a large detector solid angle and high throughput data acquisition system, offered the capacity for rapid sample analysis \cite{siddons:2014,Kirkham:2010aip}. In its construct, Maia incorporates an array of 384 silicon pin diode that are read out by twelve high energy multi-element spectrometer (HERMES) front-end ASICs and  twelve simultaneous capture of events with programmable timing and energy readout (SCEPTER) ASICs. In this two-ASIC solution, HERMES provides low noise charge preamplification and shaping \cite{DeGeronimo:2002tns} while SCEPTER performs amplitude and timing measurements for each channel along with multiplexing and event buffering \cite{oconnor:2003,Dragone:2005nss}.

The HERMES front-end ASIC is optimized in a 350 nm CMOS technology for an array of silicon pin diode sensors with pixel capacitance of $\approx$ 1 pF and leakage current of about 0.5 nA/\si{\square\centi\meter} at 27 \si{\degreeCelsius}. The noise contribution from the sensor constrained the resolution of the instrument. To better resolve low energy and closely spaced fluorescence photon lines, the Maia detector is being upgraded with new detectors, a new front-end ASIC developed in 250 nm CMOS technology and a high speed readout and processing module (RPM). 

The new detector is prototyped as two modular assemblies. The first configuration incorporates linear silicon drift detectors (SDDs). This upgrade is facilitated by a multi-amplifier readout system (MARS) ASIC prototype with selectable energy windows of 12.5 keV, 25 keV, 37.5 keV and 75 keV in silicon. The second configuration implements germanium strip detectors. These sensors are read out by a high energy MARS (HEMARS) ASIC with selectable energy windows of 25 keV, 50 keV, 100 keV and 200 keV in germanium. Each upgrade option requires either the MARS or HEMARS ASIC for charge preamplification, shaping and signal extraction. The RPM is common to both.

This paper is organized as follows. In section \ref{Sensors for Fluorescence Spectroscopy}, an overview of the linear SDD and germanium strip sensors is presented. This is followed by section \ref{MARS ASIC Architecture} with an architectural description of the MARS and HEMARS front-end ASICs. The experimental setup and the corresponding results are discussed in sections \ref{Experimental Setup} and \ref{MARS Results} respectively. We conclude with a summary of our findings and an assessment of the future work required to improve the design.

\section{Sensors for Fluorescence Spectroscopy}
\label{Sensors for Fluorescence Spectroscopy}

\subsection{Silicon Drift Detector}

\begin{figure}[!t]
    \centering\includegraphics[width=0.9\linewidth]{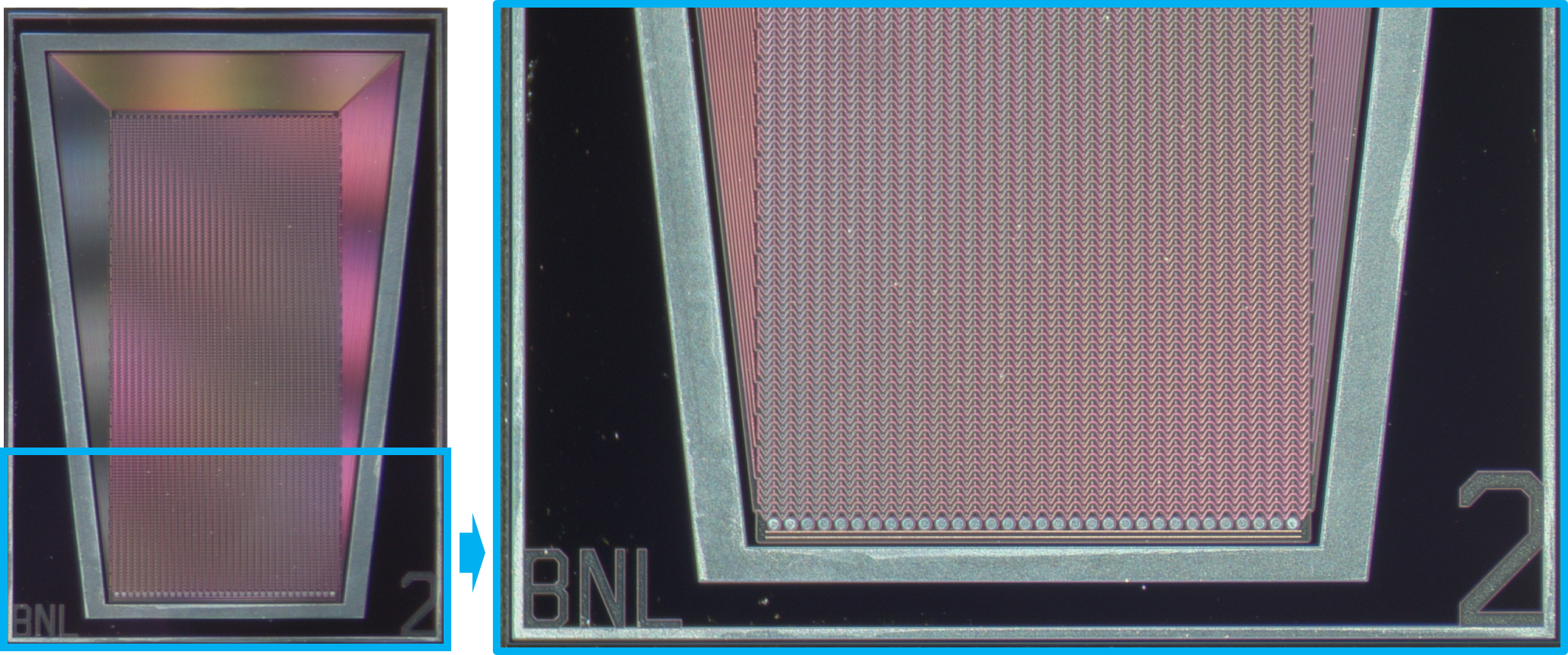}
	\caption{Photograph of the 32 anode linear SDD, with a zoom of the row of the anode pads.}
	\label{fig:sdd}
\end{figure}

At Brookhaven National Laboratory, we fabricated a multi-anode linear SDD shown in Figure \ref{fig:sdd}. SDDs are characterized by very small anode capacitance \cite{Gatti:1984nim} on the order of 50 fF. Since the anodes are very close to the edge of the chip, the wire-bonds to the MARS ASIC can be relatively short, therefore, adding a small parasitic capacitance (of approximately 100 fF). The layout of this particular linear SDD is based on the design suggested in \cite{Sonsky:1999tns}, where, on the front side,  a series of parallel sawtooth-shaped cathodes drives the electrons to the anode, while preventing the lateral diffusion of the electron cloud. The cathodes are connected to each other by means of integrated resistors, so that only the cathodes closest to and farthest from the anode row need to be biased at about -10 V and -150 V respectively. The maximum drift length of the electrons is 1 cm. Differing from the design in \cite{Sonsky:1999tns}, the back side is a uniform  p+ implant, meant to be a thin entrance window for the x-rays, as is the norm with SDDs for spectroscopic applications \cite{Lechner:2001nima,Chen:2007tns}. The entrance window is biased in the range of -60 V to -80 V, close to the depletion voltage of the substrate. The anodes are at a pitch of about 150 \si{\micro\meter}, which matches the MARS ASIC channel pitch. The sensor has an anode leakage current of about 200 pA at room temperature, dominated by the surface leakage current.

\subsection{Germanium Strip Sensor}

\begin{figure}[!t]
    \centering\includegraphics[width=1.0\linewidth]{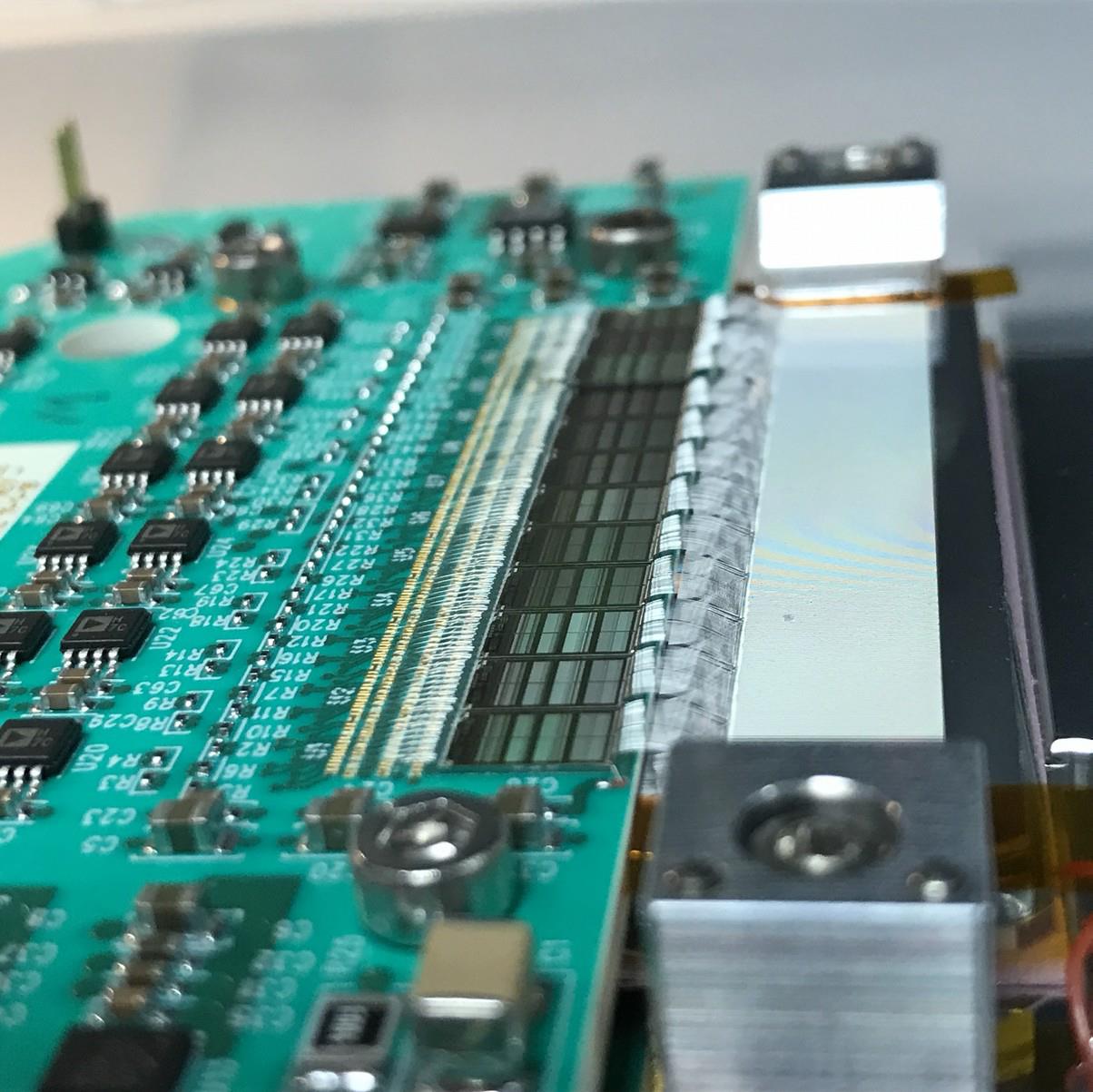}
	\caption{384- Germanium strip wire-bonded to the MARS ASIC}
	\label{fig:geNmars}
\end{figure}

We used the MARS and HEMARS ASICs for the testing and development of germanium strip detectors for a variety of synchrotron applications. Germanium has been the material of choice for high-energy x-ray and gamma ray detection. The availability of detector grade crystal is pivotal in the development of efficient detectors that can operate at or near the Fano limit. The strip sensors used in this work are fabricated at Forschungszentrum Julich \cite{krings:2011nss} and are based on a process that relies on trenching for strip isolation. These detectors are 3 mm thick and are biased just below 300 V for full depletion. The first detector system built with 12 MARS ASICs and 384 strips has a sensor geometry of 8 mm long strips separated by 30 \si{\micro\meter} wide trenches for a pitch of 0.125 mm. The sensor is mounted in thermal contact with a cold finger and insulating ceramic standoffs isolates the readout board from the cold plate. The ASICs are connected to the sensor through wire-bonds as shown in Figure \ref{fig:geNmars}. A silicon diode mounted near to the ASICs as a temperature monitor shows a temperature of about 223K while the sensor is at 100K.

\section{ASIC Architecture}
\label{MARS ASIC Architecture}
The MARS ASIC is developed to readout SDDs with adjustable energy windows from 12.5 keV to 75 keV in Silicon. The 37.5 keV and 75 keV energy windows are initially used to prototype germanium based detection systems. Subsequently, we developed a high energy multi-amplifier readout system (HEMARS) with energy settings from 25 keV to 200 keV in germanium. The architecture of the ASICs builds on the design presented in \cite{DeGeronimo:2010tns}. These new front-ends are optimized to process either positive or negative charge with continuous event driven readout. That is, each channel performs signal processing and readout independent of the others for high throughput. The informational value afforded by the architectural description that ensues applies equally to MARS and HEMARS as they differ only in topology due to the optimization of their respective signal processing chains for either silicon or germanium sensors.

\subsection{MARS and HEMARS Chip Level Architecture}

\begin{figure}[!t]
    \centering\includegraphics[width=1.0\linewidth]{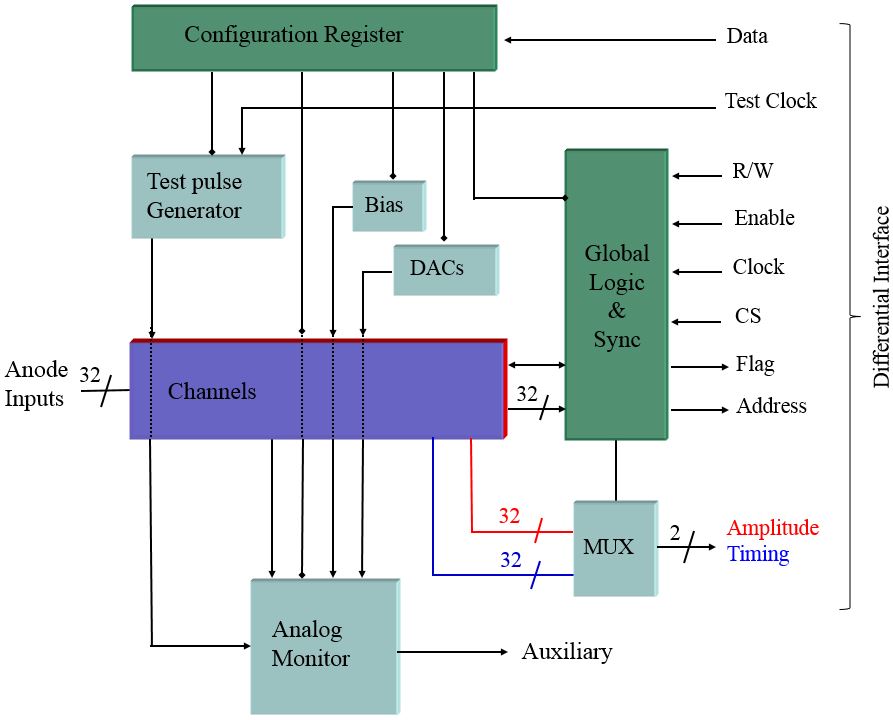}
	\caption{MARS and HEMARS block diagram with 32 input channels. Through configuration registers, the ASICs implement selectable gain and peaking times along with adjustable bias, DACs and test pulse generator. A global control logic arbitrates data write-in, acquisition, channel readout and analog monitoring.}
	\label{fig:marsBlockDiagram}
\end{figure}

A block diagram of the MARS/HEMARS ASIC is shown in Figure \ref{fig:marsBlockDiagram}. The chip comprises a configuration register, a test pulse generator, a bias, two digital-to-analog converters, thirty two input channels, an analog monitor, three 32-to-1 multiplexers, a global logic and a differential interface. 

For bench top characterization or in situ debugging of the ASIC, an embedded test pulse generator is used to inject positive or negative charge into the front-ends. An analog monitor with an auxiliary port provides the capability to verify critical analog circuit blocks. For each measurement setting, an adjustable bias network establishes the DC operating point of the circuits according to the register configuration. The back end of the chip has a global control logic which manages the ASIC during serial write-in, signal acquisition, event synchronization and readout.

The 32 input channels implement the option to process either electrons or holes with programmable gains and adjustable peaking times. When a channel with an above threshold event finds a peak, the global logic places a token in the channel and releases a flag to the external readout and processing module. The token is used to control the channel multiplexers for sparse readout of the peak amplitude, timing, corresponding address and flag status. Subsequently, the external RPM digitizes the data on the bus. If multiple channels detect a peak, they are read out sequentially from the lowest to the highest address. Since some channels are being read while others are acquiring, the ASIC interface is fully differential to minimize digital pickup. The only exception to the differential interface is the auxiliary monitor port which is used for the sole purpose of debugging.

\subsection{MARS Channel Architecture}
The MARS front-end preamplifiers are optimized for 50 fF of input capacitance and leakage currents from 100 fA to 1 nA. The input MOSFET operates at 475 \si{\micro\ampere} of drain current with a geometry (W/L) of 100.8 \si{\micro\meter}/0.36 \si{\micro\meter}  \cite{DeGeronimo:2005tns}. The primary application is the readout of silicon drift detectors that are sensitive to soft x-ray fluorescence photons. 

\begin{figure}[!t]
    \centering\includegraphics[width=1.0\linewidth]{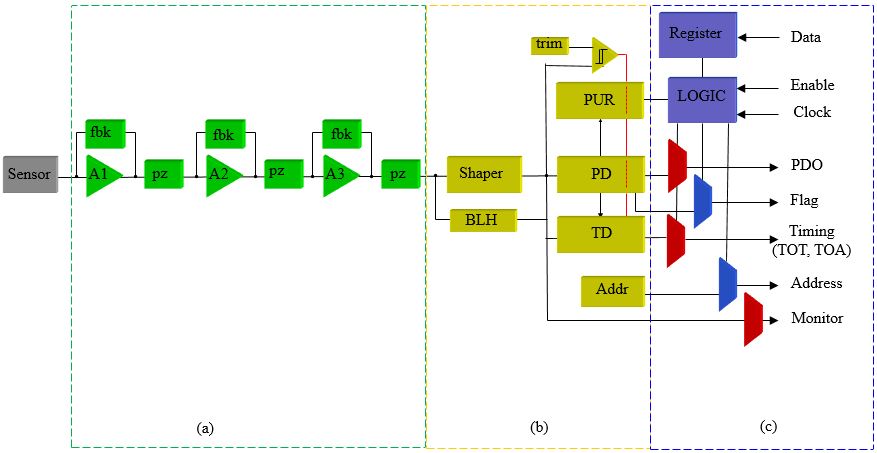}
    \caption{The MARS channel architecture illustrating (a) a multi-stage charge amplifier with adaptive feedback and polarity selection (b) a fifth order shaper with baseline holder (BLH), discriminatior with adjustable trim, peak detector (PD) and time detector (TD) for signal extraction (c) a control logic with signal multiplexers.}
	\label{fig:marsChanBlockDiagram}
\end{figure}

The channel implementation of MARS is illustrated by the block diagram in Figure \ref{fig:marsChanBlockDiagram}. The signal processing chain in each of the 32 channels is organized in three sub-blocks that perform the functions of charge preamplification, signal extraction and data multiplexing. 

The multi-stage charge amplifier is designed with three cascading stages. The first stage provides a charge gain of 38, the second charge gain stage is selectable between 14/3, 28/3, 14, and 28. The third stage serves as either a unity gain pass-through for negative charge or an inversion for positive charge to ensure signal compatibility at the shaper input. Each amplifier implements continuous adaptive reset in the feedback path followed by pole-zero cancellation \cite{Geronimo:2000tns}. By default, the channels are sensitive to negative charge but the polarity can be changed to positive charge sensitivity through a configuration register bit. 

A fifth order shaper provides analog filtering with adjustable peaking times of 0.25 \si{\micro\second}, 0.5 \si{\micro\second}, 1 \si{\micro\second} and 2 \si{\micro\second}. The shaper implementation has one real pole, while two pairs of complex conjugate poles are contributed to the constellation by two multiple feedback (MFB) stages \cite{DeGeronimo:2011tns}. Also, the shaper contributes a signal gain of $\approx$ 3.38 mV/fC to the signal for overall channel gains of 0.6 V/fC, 1.2 V/fC, 1.8 V/fC and 3.6 V/fC. Baseline stabilization is realized with a bandgap referenced baseline holder in feedback with the shaper \cite{Geronimo:2000tnsbln}. 

Thresholds are set with a 10-bit DAC and baseline dispersions are compensated by a low hysteresis discriminator with 5-bit trim per channel. The output of the discriminator serves as a trigger for above threshold events, extraction of the shaped signal amplitude with the peak detector (PD) and time over threshold (TOT) or time of arrival (TOA) measurements with the time detector (TD).

The back-end of each channel is reserved for local registers, data multiplexers, and control logic. During debugging and calibration, the configurations that are unique to the channel are managed by local registers. Similarly, since each channel operates independent of the others, the channel logic communicates with the global logic which arbitrates event acquisition and continuous event driven (CED) readout. During acquisition, the control logic ensures that all channels with events below threshold remain in acquisition mode. Further, if a channel finds a peak, the control logic processes the peak and prepares the channel for readout and then notifies the global logic of a successful event. The global logic sends a flag to the readout and processing module (RPM) which responds by sampling the peak amplitude and timing while reading the channel's address and flag status simultaneously. After the readout is completed, the channel reset then returns to acquisition mode. 

\subsection{HEMARS Channel Architecture}
The channel architecture of HEMARS is also represented by Figure \ref{fig:marsChanBlockDiagram}. The front-end channels are optimized for 50 fF of input capacitance and leakage currents from 100 fA to 1 nA. A charge gain of 30 is implemented in the first preamplifier stage while the second stage is adjustable to 7/4, 7/2, 7 and 14 respectively. The shaper contributes $\approx$ 3.51 mv/fC of gain. The overall channel gain is programmable from 184-, 368-, 736- to 1472 mV/fC corresponding to 200-, 100-, 50- and 25 keV in germanium.

\subsection{ASIC Prototype}
\begin{figure}[!t]
    \centering\includegraphics[width=0.9\linewidth]{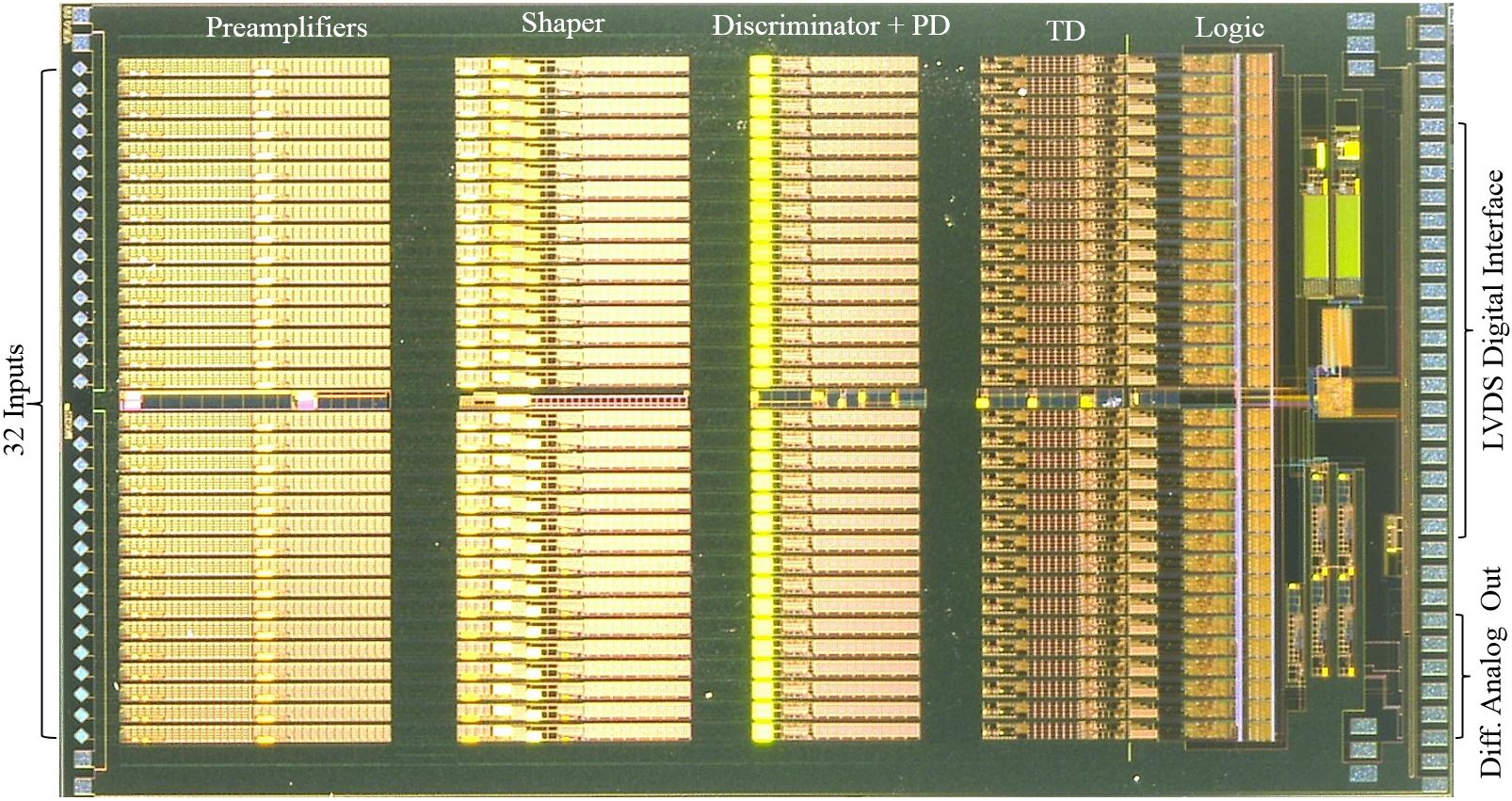}
	\caption{A micrograph of the MARS ASIC with 32 input channels. Each channel is linearly arranged as a multi-stage preamplifier, a shaper, discriminator, peak detector (PD), time detector (TD) and logic. The bias is laid out horizontally between two groups of 16 channels. The DACS, registers, and analog buffers are implemented at the back-end of chip.}
	\label{fig:marsASIClayout}
\end{figure}

A micrograph of the MARS ASIC is shown in Figure \ref{fig:marsASIClayout}. There are thirty two analog inputs that interface with the sensor. Each channel is laid out in a linear arrangement from preamplifiers, shaper, peak and time detector to local register and logic. At the back-end of the chip are the digital-to-analog converters (DACs), the global configuration registers and global logic. The low-voltage differential signaling (LVDS) pads for the digital inputs and outputs are located on the top right edge of the chip and isolated from the analog pads to prevent coupling. No pads are allowed on the horizontal sides of the ASIC as the chips are abutted to match the sensor pixel pitch in order to minimize spreading of the wire-bonds. The latter constrains the maximum die cut size to 6.3 mm x 3.9 mm. Twelve MARS ASICs are required to read out the 384 anodes of the Maia detector. 

\section{Electronic Hardware Integration}
\label{Experimental Setup}

\begin{figure}[!t]
    \centering\includegraphics[width=1.0\linewidth]{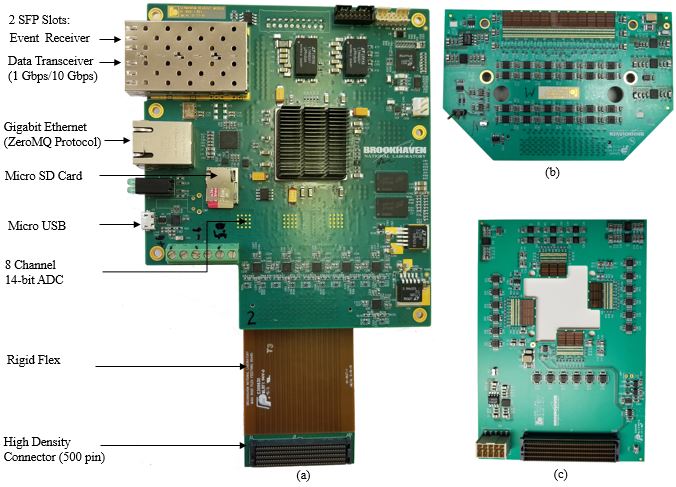}
	\caption{Maia detector upgrade electronics. (a) External readout and processing module (RPM) with Zilinx Zynq-7000 SoC FPGA including embedded ARM Cortex A9 processor. Two 1 Gbps SFP slots for event receiver and data transmission, Giga-bit Ethernet and 3 eight channel 25 MSPS 14-bit ADC. High density rigid-flex cable with 500 pin connector (b) HEMARS carrier board with 12 ASICs mounted in a linear arrangement. (c) MARS carrier board with 3 ASICs mounted per quadrant, compatible to the Maia detector.}
	\label{fig:marsDAQ}
\end{figure}

The full signal processing hardware shown in Figure \ref{fig:marsDAQ} is comprised of a readout and processing module (RPM) that interfaces with either the HEMARS carrier board or the MARS carier board through a rigid flex interconnect printed circuit board. The assembled system is used for ASIC characterization, x-ray fluorescence spectroscopy and energy dispersive x-ray diffraction.

The RPM in Figure \ref{fig:marsDAQ}a includes a Xilinx\textsuperscript{\textregistered} Zynq-7000 SoC FPGA with dual ARM Cortex\textsuperscript{\texttrademark}-A9 processor running Debian Linux. The FPGA bitfile, linux kernel, and root file system are stored on a micro SD card. For high speed data, two small form-factor pluggable (SFP) transceiver slots are implemented. One slot is for an embedded event receiver that is compatible with the timing system at the National Synchrotron Light Source II (NSLS II) at Brookhaven National Laboratory. The other SFP slot uses the UDP communication protocol for  transmission of list mode event data from the detector over Gigabit Ethernet.  Further, the user interface is accessible either through Gigabit Ethernet connection (ZeroMQ Protocol) or console login via the micro USB serial port. Three 8-channel ADCs with 14-bit resolution are mounted on the back of Figure \ref{fig:marsDAQ}a. The ADCs sample analog data from 12 ASICs, each with a dedicated amplitude (PD) and timing (TD) port. With the ADC running continuously at 25 Mbps, the data throughput is $\approx$ 8.4 Gbps (3 x 8 x 14 x 25 ) without any meta data headers. Detailed information of the RPM is reported in \cite{Rumaiz:2018jinst}.

The ASIC carrier boards are fabricated from low loss Rodgers RO4000\textsuperscript{\textregistered} laminates and populated with discrete components that are vacuum compatible at \num{e-6} torr. On each carrier board, the ASIC locations are determined by the sensor geometry. The HEMARS carrier board with 12 ASICs installed in a linear arrangement without the sensor is shown in Figure \ref{fig:marsDAQ}b. Figure \ref{fig:geNmars} presents a picture of the same module with the Ge strip sensor wire-bonded to the ASICs. Similarly, the MARS carrier board with 3 ASICs mounted on each quadrant without the SDD is shown in Figure \ref{fig:marsDAQ}c. The other discrete components are identical for both boards. Most notable on each carrier are the 500 pin high density connectors, 24 buffers (2 per ASIC) that drive the external ADCs on the RPM and the low dropout regulators that supply power to the ASICs. Each detector module is capable of reading out a maximum of 384 Ge strips or 384 SDD anodes.


\section{Results}
\label{MARS Results}

\begin{figure}[!t]
    \centering\includegraphics[width=1.0\linewidth]{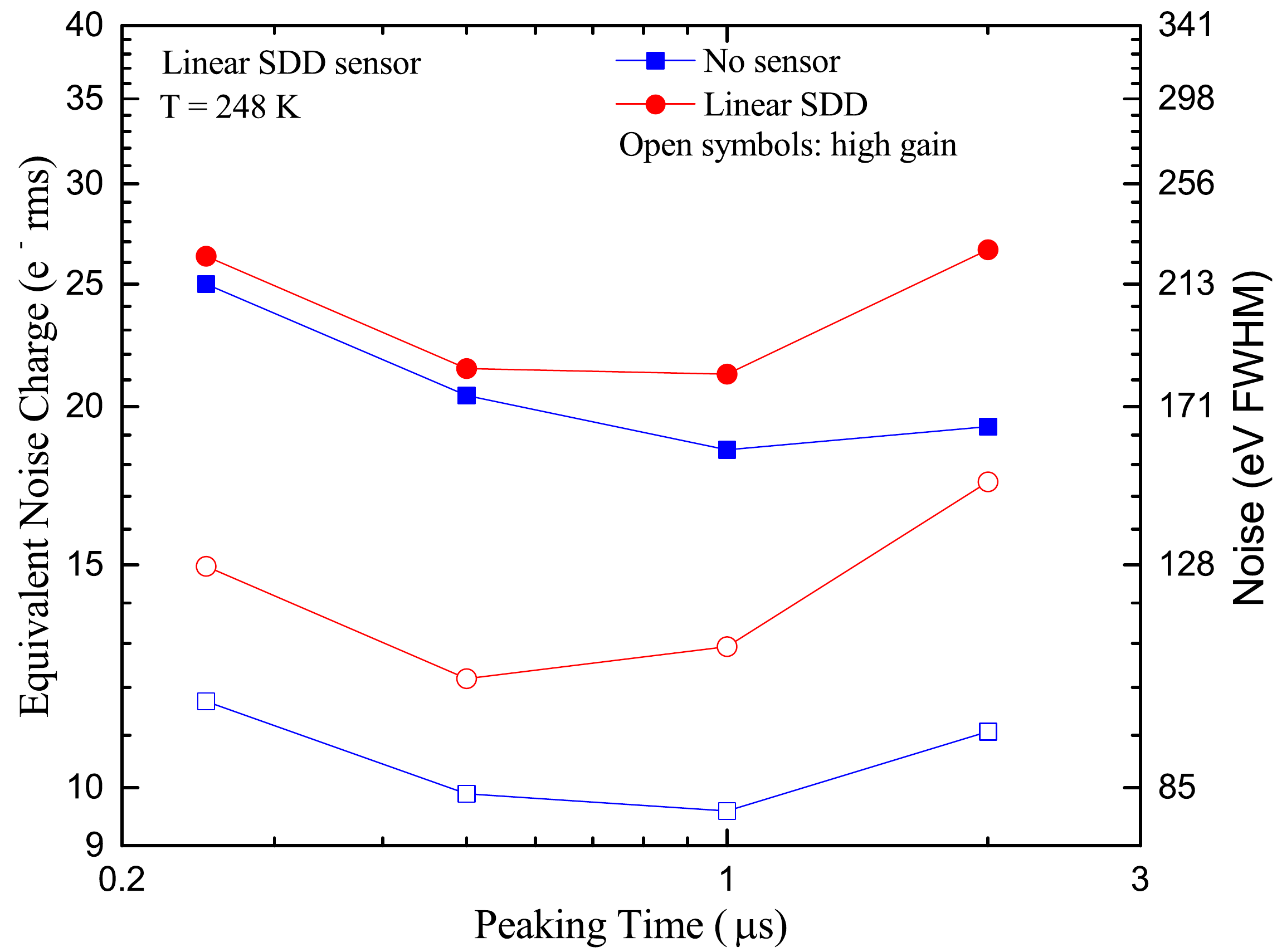}
	\caption{MARS ENC as a function of peaking time without sensor (blue trace) and with the linear SDD (red trace) at 248K at the shaper output. The solid symbols are for measurements at 600 mV/fC (75 keV) while the open symbols are measurements taken at 3.6 V/fC (12.5 keV).}
	\label{fig:marsEnc}
\end{figure}

\subsection{Results from the MARS ASIC Prototype}
A true rms meter was used to measure the noise voltage at the output of the shaper on the auxiliary monitoring port. Figure \ref{fig:marsEnc} shows the equivalent noise charge (ENC) as a function of peaking time with (red trace) and without (blue trace) the linear SDD connected at 248 K. At a gain of 3600 mV/fC (12.5 keV energy range in Si) and 1 \si{\micro\second} peaking time, an ENC of 10 \si{\electron} rms was obtained which corresponds to a dynamic range (DR) of $\approx$ 147. With the sensor connected, the ENC increased to 13 electrons (DR of $\approx$ 113). At low gain (600 mV/fC corresponding to 75 keV) and 1 \si{\micro\second} peaking time, we measured 21- and 18 \si{\electron} rms with and without the sensor, corresponding to a dynamic range of 488 and 419 respectively. 

The increase in ENC at each gain setting with the sensor connected was due to the added parasitic interconnection capacitance, the sensor anode capacitance ($\approx$ 100 fF) plus the sensor leakage current. Regarding the difference in total ENC between low gain (600 mV/fC) and high gain (3600 mV/fC), this contribution came from the shaper. Compared to the front-end in \cite{DeGeronimo:2002tns}, the electronic noise without sensor decreased from 14 \si{\electron} rms to 10 \si{\electron} rms in MARS. This is an improvement for the Maia upgrade. In addition, the front-end power dissipation was reduced from 11 mW/ch to 3.6 mW/ch. At the system integration level, the assembly process was simplified as the HERMES and SCEPTER pair were replaced by a one ASIC solution. That is, either MARS or HEMARS  performed complete readout of the sensor, since each channel implemented the full signal processing chain with dedicated peak and time detectors. 

\begin{figure}[!t]
    \centering\includegraphics[width=1.0\linewidth]{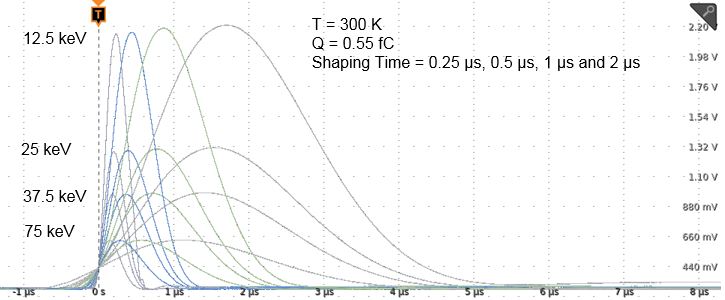}
	\caption{MARS shaper response to 0.55 fC of injected charge at each energy setting and peaking time.}
	\label{fig:marsShResponse}
\end{figure}

The integrated test generator was used to inject 0.55 fC of charge into the front-end. The chip was configured so that the shaper output of a channel was multiplexed onto the auxiliary port where it was monitored with an oscilloscope. Figure \ref{fig:marsShResponse} shows the recorded waveform at each gain and peaking time. At 3.6 V/fC, the semi-Gaussian pulse responses had a maximum amplitude of approximately 2.2 V and a nearly symmetrical return to baseline.


\begin{figure}[!t]
    \centering\includegraphics[width=1.0\linewidth]{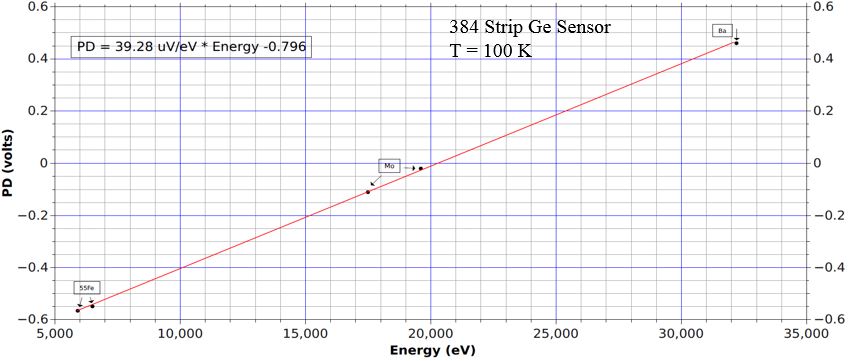}
	\caption{Energy calibration of detector with a 384 strip Ge sensor at 100K read out by 12 MARS ASICs (1.2 V/fC). X-ray photons were from $^{55}$Fe source and molybdenum and barium targets excited by $^{241}$Am. The linearity was better than $\pm$ 1 \si{\percent}}
	\label{fig:linearity}
\end{figure}

At a gain setting of 1.2 mV/fC and 1 \si{\micro\second} peaking time, the MARS ASIC was used to readout a biased 384 strip germanium sensor that was cooled to  100 K. An $^{55}$Fe source and an $^{241}$Am source used to excite molybdenum (Mo) and barium (Ba) targets were used to determine the ASIC linearity over the 37.5 keV energy range. From left to right, Figure \ref{fig:linearity} shows the manganese (Mn) K$_{\alpha}$ (5.9 keV) and K$_{\beta}$ (6.5 keV), Mo K$_{\alpha}$ (17.4 keV) and K$_{\beta}$ (19.6 keV) and Ba K$_{\alpha}$ (32.2 keV) characteristic x-ray photo peaks. A linearity better than $\pm$ 1\si{\percent} was achieved.

\begin{figure*}[!t]
    \centering\includegraphics[width=1\linewidth]{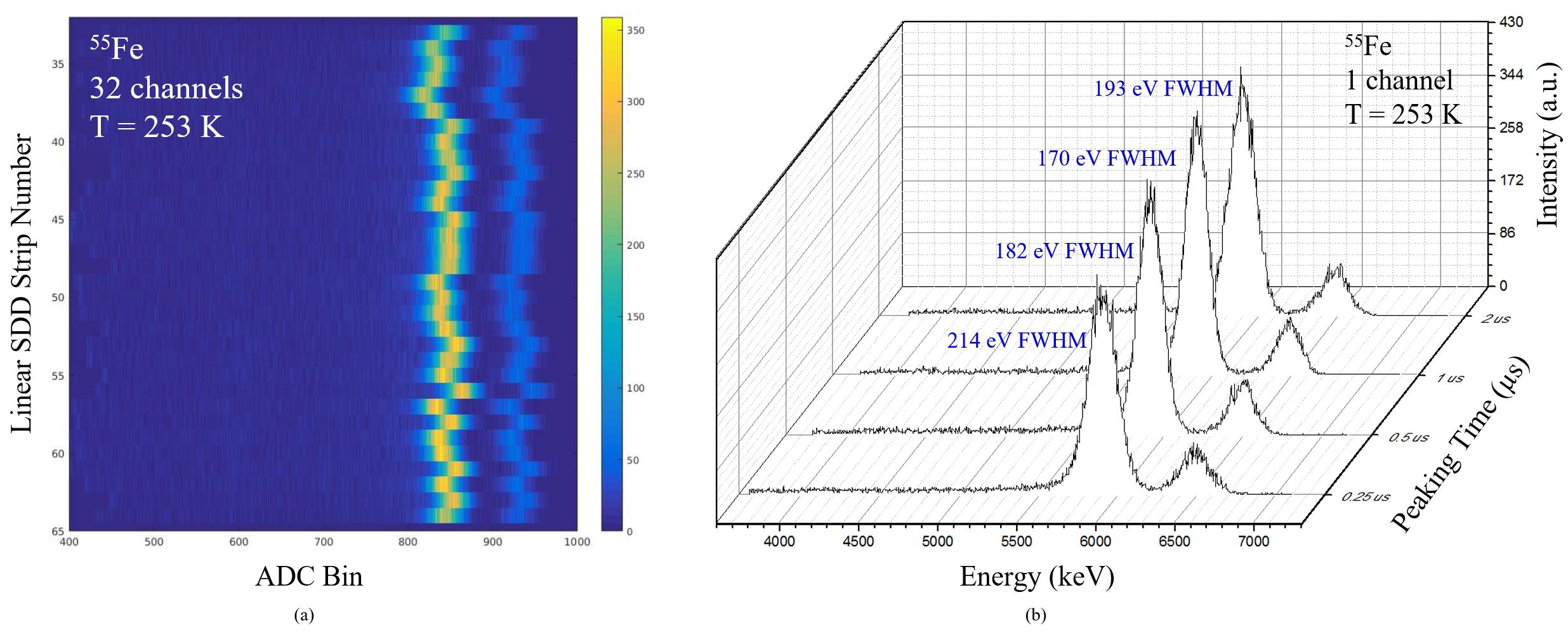} 
	\caption{MARS ASIC readout of a linear SDD. (a) Intensity map from a 32 anode linear SDD  with the Mn K$_{\alpha}$ and K$_{\beta}$ lines. (b) Spectral resolution of the Mn K$_{\alpha}$ peak range from 170 - 214 eV FWHM from 0.25 - 2 \si{\micro\second} peaking time.}
	\label{fig:marsFWHM}
\end{figure*}

In addition, Spectral measurements with an $^{55}$Fe source were taken with a 32 anode linear SDD cooled to 253 K. The MARS ASIC was configured for a gain of 3.6 V/fC (12.5 KeV energy range) and the acquisition was executed at each peaking time. Figure \ref{fig:marsFWHM}a shows the intensity map of the 32 ASIC channel response. The Mn K$_{\alpha}$ (5.9 keV) and K$_{\beta}$ (6.5 keV) lines are clearly seen. The fitted spectra from one channel is shown in Figure \ref{fig:marsFWHM}b. The measured FWHM at 5.9 keV were 214 eV, 182 eV, 170 eV and 193 eV at 0.25 \si{\micro\second}, 0.5 \si{\micro\second}, 1 \si{\micro\second} and 2 \si{\micro\second} peaking time respectively. These results confirmed the ENC measurement with the linear SDD at high gain in Figure \ref{fig:marsEnc}. Without a radiation source in Figure \ref{fig:marsEnc}, the total noise amounted to the measured contributions from the readout electronics, sensor leakage current and capacitance. For the spectra in Figure \ref{fig:marsFWHM}b, signal formed inside the sensor experienced statistical fluctuations that contributed to the noise according to Fano statistics. Inevitably, the photo-peak was broadened with an effective resolution given by $\sigma_{eff} = \sqrt{ {\sigma_{stat}}^{2} + ENC^{2} }$.

\begin{figure}[!t]
    \centering\includegraphics[width=1.0\linewidth]{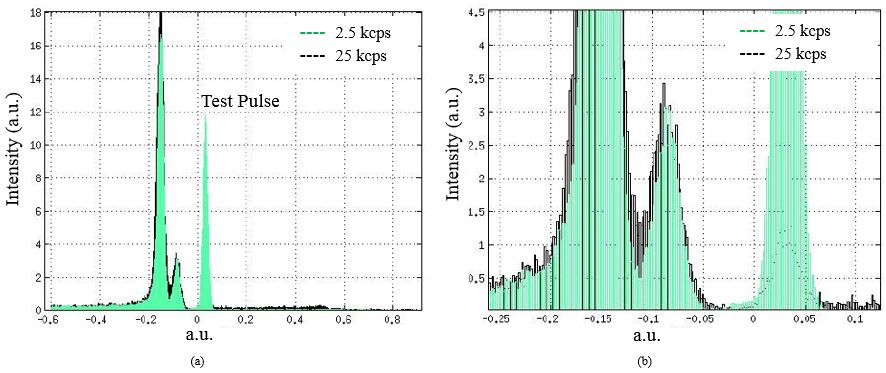}
	\caption{MARS ASIC readout of a 96 pixel silicon PIN diode sensor cooled to -20 \si{\degreeCelsius}.(a) An $^{55}$Fe source and internal test pulse rate of 2.5 kcps (green) and 25 kcps (black). (b) Zoomed in segment shows a photopeak and test pulse shifts of 2 a.u. bins.}
	\label{fig:rate}
\end{figure}

We investigated the impact of event rate on the photopeak location of spectra collected with the MARS ASIC reading out a 96 pixel silicon PIN diode sensor cooled to -20 \si{\degreeCelsius}. Both an $^{55}$Fe source and the internal test pulse were tuned for event rates of 2.5 kHz and 25 kHz. From the plots in Figures \ref{fig:rate}a and \ref{fig:rate}b, the peak shift was approximately 2 arbitrary unit (a.u.) bins as evidenced by the centroids of the Mn K$_{\alpha}$ and K$_{\beta}$ photopeaks and the test pulse. 

\subsection{Results from the HEMARS ASIC}

\begin{figure}[!t]
    \centering\includegraphics[width=1.0\linewidth]{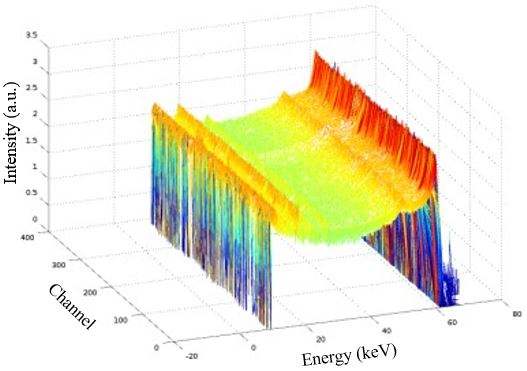}
	\caption{Spectral response of 12 HEMARS ASICs reading out 384 strips of Ge sensor irradiated by $^{241}$Am source.}
	\label{fig:geNmarsRes}
\end{figure}

In diffraction studies, photons with energies up to 200 keV are of interest. Since the MARS chip was primarily designed for use with silicon based detectors, the gain setting was limited to 75 keV. Analysis with germanium detectors was geared to be used in high energy diffraction experiments with photon energy as high as 200 keV. The HEMARS ASIC addressed the latter as its gain selections were optimized for 25 keV to 200 keV in Ge. Figure \ref{fig:geNmarsRes} shows the response of all 384 channels to an $^{241}$Am radioactive source. This has a strong line at 59.5 KeV and several weak lines at lower energies. The measured resolution was 610 eV FWHM at 59.5 keV. 

\begin{figure}[!t]
    \centering\includegraphics[width=1.0\linewidth]{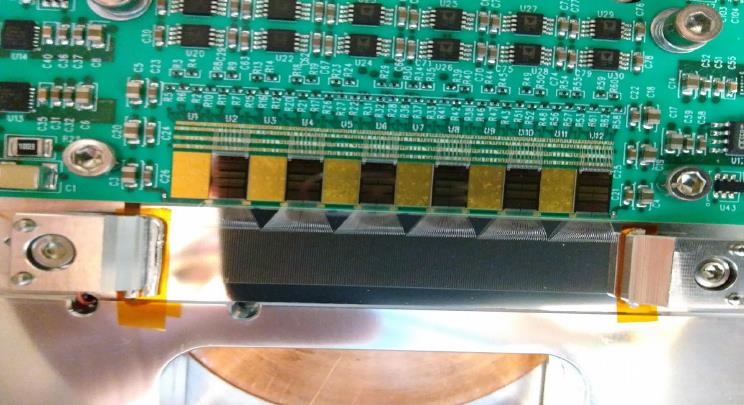}
	\caption{Completed 192-strip detector with HE-MARS. The gold pad shows the omitted ASICs}
	\label{fig:geNhemars}
\end{figure}

A 192-strip detector with 0.25 mm pitch was interfaced with the HE-MARS chip. The same ASIC carrier board as the 384-strip was used, except to accommodate the 192-strip, alternate ASICs were omitted. Figure \ref{fig:geNhemars} shows the detector assembly. The detector showed an energy resolution of about 770 eV FWHM for a 122 KeV line from a $^{57}$Co source. Details of this results can be found in \cite{Rumaiz:2018jinst}.

\begin{figure}[!t]
    \centering\includegraphics[width=1.0\linewidth]{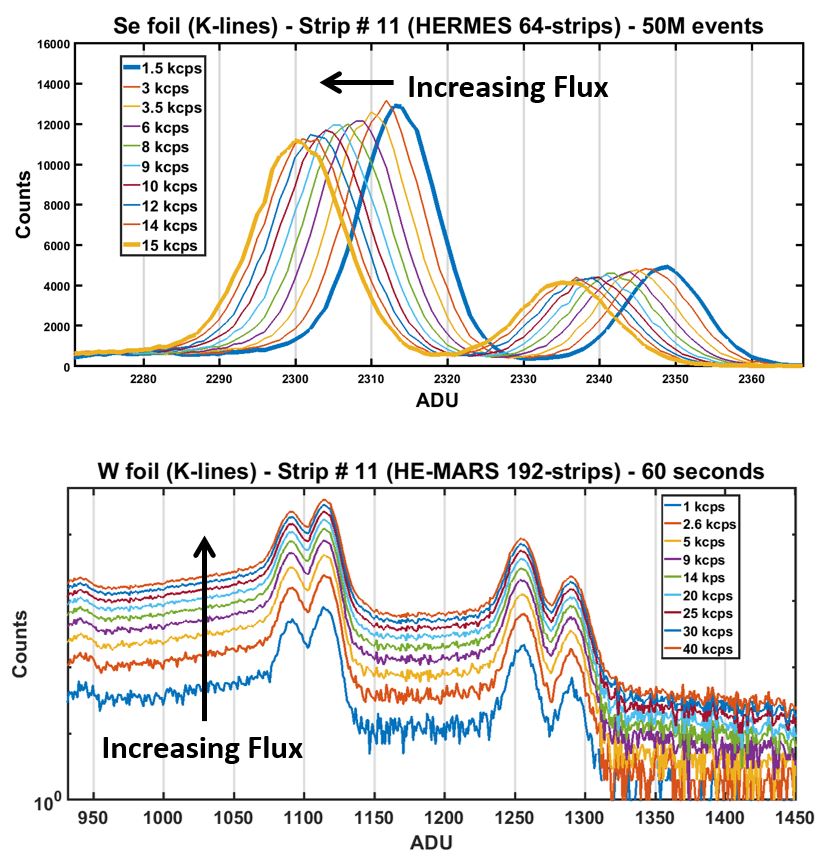}
	\caption{(a) HERMES x-ray spectra of selenium with a peak shift of approximately one pulse width when the flux increased from 1.5 kcps to 15 kcps. (b) HEMARS spectra of tungsten with a peak shift of less than 2 analog-to-digital unit (ADU) bins for flux up to 40 kcps.}
	\label{fig:peakshift}
\end{figure}

Figure \ref{fig:peakshift} shows the measured effect of event rate on peak position with HERMES ASIC and HEMARS prototype. Selenium (Se) and tungsten (W) foils were excited with the full white beam spectrum at the Argonne National Laboratory Advance Photon Science (APS) bending magnet 1-BM. This beamline contains photon from 5 keV to energies beyond 100 keV \cite{Lang:1999rsi}. Germanium strip sensors cooled to 100 K were used to detect the secondary x-ray photons from the targets. The Se foil spectra acquired with two HERMES ASICs reading out a 64 strip germanium sensor shows the Se K$_{\alpha}$ (11.22 keV) and K$_{\beta}$ (12.49 keV) lines respectively.  For this readout system, the photopeak shifted by almost one pulse width from 1.5 kcps to 15 kcps. Similar measurements were taken with fluorescence photons from a W target. Here, a 192 strip detector was readout by 6 HEMARS ASICs. The tungsten spectra shows the K$_{\alpha1}$ (59.32 keV), K$_{\alpha2}$ (57.98 keV), K$_{\beta1}$ (67.24 keV) and K$_{\beta2}$ (69.08 keV) x-ray emission lines respectively. For rates from 1kcps to 40 kcps, the peak shift was about 2 analog-to-digital unit (ADU) bins. The added capability of high rate and high throughput was another upgrade to the Maia detector \cite{siddons:2014}.   

\begin{figure}[!t]
    \centering\includegraphics[width=1.0\linewidth]{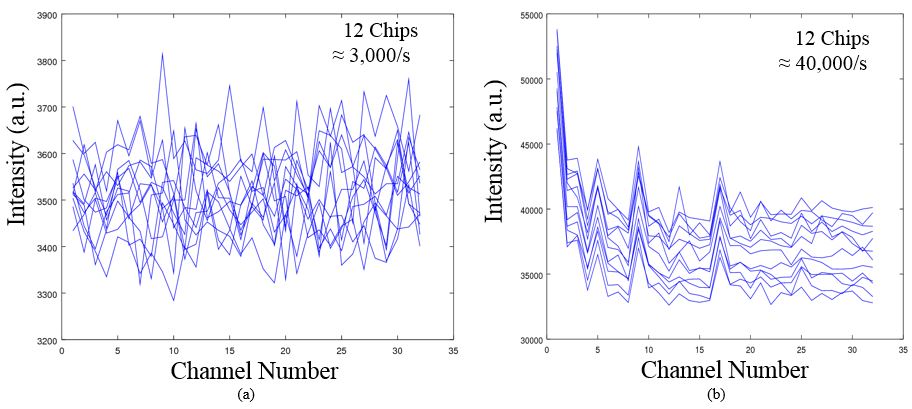}
	\caption{ 12 HEMARS ASIC event response to a flood field illumination from a 17.45 keV x-ray source. (a) at a rate of $\approx$ 3000/s, the integrated intensity shows normal statistical fluctuations. (b) at a rate of 40,000/s, the low address channels are systematically processed with preference above high order channels.}
	\label{fig:floodField}
\end{figure}

The MARS and HEMARS ASICs demonstrated minimal shift in the photopeak with rate. However, it was observed that the token passing readout scheme exhibited preferential bias towards low address channels at high event rates. We used an x-ray generator producing 17.4 keV photons (Mo K$_{\alpha}$) to uniformly irradiate a 384 strip Ge sensor at 100 K. The intensity on each channel of the HEMARS ASIC was captured in Figure  \ref{fig:floodField}. Each line represents the count from 32 channels on one ASIC. At a rate of 3,000 events/s, the 12 chips reported the expected integrated intensity shown in Figure \ref{fig:floodField}a, with the anticipated statistical fluctuations. However, when the rate was increased to 40,000 events/s, it was observed that the integrated intensity of low address channels was higher than those with higher address. The periodic spikes are explained by the algorithm used in the logic to arbitrate clusters of four channels. Even though the ASIC implemented a sparse readout, the token entered the chip at channel 1 and exited at channel 32. As a result, at high rates, the low order channels experienced preferential access to the RPM. Ideally, it was desired for channels to be read in the order in which events arrived. To explain Figure \ref{fig:floodField}b, consider a high rate scenario in which channel 20 acquired an event and signaled to the ASIC global logic that it was ready to be read. The global logic responded by releasing a token which should bypass all channels and stop in channel 20. Now, assume that channel 6 released a flag within a clock cycle of 20 reporting, the token would be captured by channel 6. The problem was further exacerbated if during the time the RPM takes to digitize the analog outputs of channel 6, other channels with addresses greater than 6 and less than 20 acquired events. This put channel 20 further in the queue to be read. This issue will be addressed in the next revision to prioritize readout based on the order in which the channels report to the global logic. 

\section{Conclusions and Future Work}
\label{ MARS Conclusions and Future Work}

Compared to the HERMES ASIC, the MARS and HEMARS ASIC prototypes provided improved noise, rate and power performance for the Maia detector upgrade. Although there were enhancements, the ASIC rate was impacted by the 200 ns settling time required for the chip to come out of analog tri-state. Further, at rates around 40 kcps and above, the token passing scheme showed preferential readout to low address channels. The ASICs have demonstrated the capability to readout both silicon and germanium sensors with high resolutions for XFS and EDX experiments at synchrotrons, just to name a few applications.  

In our characterization of the ASICs, two issues were observed that will be addressed in the next revision. First, the initial concept for the system level design required that groups of 3 ASICs share a common ADC on the RPM. This requirement was met at the ASIC level through the implementation of analog tri-states on the peak detector output (PD) and the time detector output (TD) so that only the chip being read could put data on the bus. It was observed that whenever a chip was first accessed for readout, as it came out of tri-state, PD and TD were slow to settle. On average the settling time should be about 10 ns but TD took as long as 200 ns to settle on the first read, after which it behaves normally. The RPM upgrade implements parallel readout of the ASICs. Therefore, the tri-stating capability is no longer required and will be removed. Another improvement involves optimizing the signal path from the peak and time detectors to the external ADC. That is, lowering the parasitic load while increasing the driving capability of the buffer. In the next prototype, the on-chip buffer will be revised to drive the external ADC directly. 

Second, the issue with the token favoring low address channels at high rates will be addressed by eliminating the token passing scheme. Instead, a FIFO based logic will be used to record the order in which the channels acquire events and perform the readout in the same manner.


%



\section*{Acknowledgment}
The authors are grateful to our collaborators from the National Synchrotron Light Source II and the Instrumentation Division at Brookhaven National Laboratory. Special thanks to our colleagues from the X-ray Science Division at Argonne National Laboratory.

We are also grateful to Dr. Eliane Lessner for her support. This work was supported by the Accelerator and Detector R\&D Program in Basic Energy Sciences (BES), Scientific User Facilities (SUF) Division at the Department of Energy. 

This research uses resources of the National Synchrotron Light Source II, a U.S. Department of Energy (DOE) Office of Science User Facility operated for the DOE Office of Science by Brookhaven National Laboratory under Contract No. DE-SC0012704. BNL is supported by the U.S. Department of Energy, Office of Science, Office of Basic Energy Sciences under contract No. DE-AC02-98CH10886.

In addition, this research uses the resources of the Advanced Photon Source, a U.S. Department of Energy (DOE) Office of Science User Facility operated for the DOE Office of Science by Argonne National Laboratory under Contract No. DE-AC02-06CH11357.

\ifCLASSOPTIONcaptionsoff
  \newpage
\fi



%


\bibliographystyle{IEEEtran}
\bibliography{bibtex/bib/IEEEabrv.bib,bibtex/bib/IEEEexample.bib}{}


%








\end{document}